\documentstyle[twoside,psfig]{article}

\catcode`\@=11
\long\def\@makefntext#1{
\protect\noindent \hbox to 3.2pt {\hskip-.9pt  
$^{{\eightrm\@thefnmark}}$\hfil}#1\hfill}		

\def\@makefnmark{\hbox to 0pt{$^{\@thefnmark}$\hss}}	
	
\def\ps@myheadings{\let\@mkboth\@gobbletwo
\def\@oddhead{\hbox{}
\rightmark\hfil\eightrm\thepage}   
\def\@oddfoot{}\def\@evenhead{\eightrm\thepage\hfil
\leftmark\hbox{}}\def\@evenfoot{}
\def\sectionmark##1{}\def\subsectionmark##1{}}

\oddsidemargin=\evensidemargin
\newcounter{sectionc}\newcounter{subsectionc}\newcounter{subsubsectionc}
\renewcommand{\section}[1] {\vspace{12pt}\addtocounter{sectionc}{1} 
\setcounter{subsectionc}{0}\setcounter{subsubsectionc}{0}\noindent 
	{\tenbf\thesectionc. #1}\par\vspace{5pt}}
\renewcommand{\subsection}[1] {\vspace{12pt}\addtocounter{subsectionc}{1} 
	\setcounter{subsubsectionc}{0}\noindent 
	{\bf\thesectionc.\thesubsectionc. {\kern1pt \bfit #1}}\par\vspace{5pt}}
\renewcommand{\subsubsection}[1] {\vspace{12pt}\addtocounter{subsubsectionc}{1}
	\noindent{\tenrm\thesectionc.\thesubsectionc.\thesubsubsectionc.
	{\kern1pt \tenit #1}}\par\vspace{5pt}}
\newcommand{\nonumsection}[1] {\vspace{12pt}\noindent{\tenbf #1}
	\par\vspace{5pt}}

\newcounter{appendixc}
\newcounter{subappendixc}[appendixc]
\newcounter{subsubappendixc}[subappendixc]
\renewcommand{\thesubappendixc}{\Alph{appendixc}.\arabic{subappendixc}}
\renewcommand{\thesubsubappendixc}
	{\Alph{appendixc}.\arabic{subappendixc}.\arabic{subsubappendixc}}

\renewcommand{\appendix}[1] {\vspace{12pt}
        \refstepcounter{appendixc}
        \setcounter{figure}{0}
        \setcounter{table}{0}
        \setcounter{lemma}{0}
        \setcounter{theorem}{0}
        \setcounter{corollary}{0}
        \setcounter{definition}{0}
        \setcounter{equation}{0}
        \renewcommand{\thefigure}{\Alph{appendixc}.\arabic{figure}}
        \renewcommand{\thetable}{\Alph{appendixc}.\arabic{table}}
        \renewcommand{\theappendixc}{\Alph{appendixc}}
        \renewcommand{\thelemma}{\Alph{appendixc}.\arabic{lemma}}
        \renewcommand{\thetheorem}{\Alph{appendixc}.\arabic{theorem}}
        \renewcommand{\thedefinition}{\Alph{appendixc}.\arabic{definition}}
        \renewcommand{\thecorollary}{\Alph{appendixc}.\arabic{corollary}}
        \renewcommand{\theequation}{\Alph{appendixc}.\arabic{equation}}
        \noindent{\tenbf Appendix \theappendixc #1}\par\vspace{5pt}}
\newcommand{\subappendix}[1] {\vspace{12pt}
        \refstepcounter{subappendixc}
        \noindent{\bf Appendix \thesubappendixc. {\kern1pt \bfit #1}}
	\par\vspace{5pt}}
\newcommand{\subsubappendix}[1] {\vspace{12pt}
        \refstepcounter{subsubappendixc}
        \noindent{\rm Appendix \thesubsubappendixc. {\kern1pt \tenit #1}}
	\par\vspace{5pt}}

\newcommand{\textlineskip}{\baselineskip=13pt}
\newcommand{\smalllineskip}{\baselineskip=10pt}

\def\eightcirc{
\begin{picture}(0,0)
\put(4.4,1.8){\circle{6.5}}
\end{picture}}
\def\eightcopyright{\eightcirc\kern2.7pt\hbox{\eightrm c}} 

\newcommand{\copyrightheading}[1]
	{\vspace*{-2.5cm}\smalllineskip{\flushleft
	{\footnotesize International Journal of Modern Physics A #1}\\
	{\footnotesize $\eightcopyright$\, World Scientific Publishing
	 Company}\\
	 }}

\newcommand{\publisher}[2]{{\begin{center}\footnotesize\smalllineskip 
	Received #1\\
	Revised #2
	\end{center}
	}}

\def\abstracts#1#2#3{{
	\centering{\begin{minipage}{4.5in}\footnotesize\baselineskip=10pt
	\parindent=0pt #1\par 
	\parindent=15pt #2\par
	\parindent=15pt #3
	\end{minipage}}\par}} 

\def\keywords#1{{
	\centering{\begin{minipage}{4.5in}\footnotesize\baselineskip=10pt
	{\footnotesize\it Keywords}\/: #1
	 \end{minipage}}\par}}

\renewenvironment{thebibliography}[1]
	{\frenchspacing
	 \ninerm\baselineskip=11pt
	 \begin{list}{\arabic{enumi}.}
	{\usecounter{enumi}\setlength{\parsep}{0pt}
	 \setlength{\leftmargin 12.7pt}{\rightmargin 0pt} 
	 \setlength{\itemsep}{0pt} \settowidth
	{\labelwidth}{#1.}\sloppy}}{\end{list}}

\newcounter{itemlistc}
\newcounter{romanlistc}
\newcounter{alphlistc}
\newcounter{arabiclistc}

    	{\setcounter{itemlistc}{0}		
	 \begin{list}{$\bullet$}		
	{\usecounter{itemlistc}			
	 \leftmargin10pt		
	 \setlength{\parsep}{0pt}
	 \setlength{\itemsep}{0pt}	
	}}{\end{list}}

	{\setcounter{romanlistc}{0}		
	 \begin{list}{$($\roman{romanlistc}$)$}	
	{\usecounter{romanlistc}		
	 \leftmargin18pt 
	 \setlength{\parsep}{0pt}
	 \setlength{\itemsep}{0pt}	
	 \settowidth{\labelwidth}{#1}                          
	}}{\end{list}}

	{\setcounter{enumii}{0}			
	 \begin{list}{$($\alph{enumii}$)$}	
	{\usecounter{enumii}			
	 \leftmargin18pt		
	 \setlength{\parsep}{0pt}
	 \setlength{\itemsep}{0pt}	
	 \settowidth{\labelwidth}{#1}                          
	}}{\end{list}}

\newcommand{\fcaption}[1]{
        \refstepcounter{figure}
        \setbox\@tempboxa = \hbox{\footnotesize Fig.~\thefigure. #1}
        \ifdim \wd\@tempboxa > 5in
           {\begin{center}
        \parbox{5in}{\footnotesize\smalllineskip Fig.~\thefigure. #1}
            \end{center}}
        \else
             {\begin{center}
             {\footnotesize Fig.~\thefigure. #1}
              \end{center}}
        \fi}

\newcommand{\tcaption}[1]{
        \refstepcounter{table}
        \setbox\@tempboxa = \hbox{\footnotesize Table~\thetable. #1}
        \ifdim \wd\@tempboxa > 5in
           {\begin{center}
        \parbox{5in}{\footnotesize\smalllineskip Table~\thetable. #1}
            \end{center}}
        \else
             {\begin{center}
             {\footnotesize Table~\thetable. #1}
              \end{center}}
        \fi}

\def\tablefont{\footnotesize}
\def\tablecaptionfont{\footnotesize}
\long\def\tbl#1#2{%
	\parindent\z@\ignorespaces\noindent\tablecaptionfont
	\caption{#1}%
  	\par\setbox\tempbox\hbox{\tablefont #2}%
  	\tablewd\hsize\advance\tablewd-\wd\tempbox\global\divide\tablewd\tw@
	\ifdim\wd\captionbox<\wd\tempbox\centerline{\unhbox\captionbox}
	\else\leftskip\tablewd\rightskip\leftskip{\unhbox\captionbox}\par
	\fi\vskip5pt\centerline{\box\tempbox}
}%

\def\@citex[#1]#2{\if@filesw\immediate\write\@auxout
	{\string\citation{#2}}\fi
\def\@citea{}\@cite{\@for\@citeb:=#2\do
	{\@citea\def\@citea{,}\@ifundefined
	{b@\@citeb}{{\bf ?}\@warning
	{Citation `\@citeb' on page \thepage \space undefined}}
	{\csname b@\@citeb\endcsname}}}{#1}}

\newif\if@cghi
\def\cite{\@cghitrue\@ifnextchar [{\@tempswatrue
	\@citex}{\@tempswafalse\@citex[]}}
\def\citelow{\@cghifalse\@ifnextchar [{\@tempswatrue
	\@citex}{\@tempswafalse\@citex[]}}
\def\@cite#1#2{{$\null^{#1}$\if@tempswa\typeout
	{IJCGA warning: optional citation argument 
	ignored: `#2'} \fi}}

\def\pmb#1{\setbox0=\hbox{#1}
	\kern-.025em\copy0\kern-\wd0
	\kern.05em\copy0\kern-\wd0
	\kern-.025em\raise.0433em\box0}

\def\fnt#1#2{\footnotetext{\kern-.3em
	{$^{\mbox{\scriptsize #1}}$}{#2}}}

\def\@makefnmark{\hbox to 0pt{$^{\@thefnmark}$\hss}}	
	
\def\ps@myheadings{%
    \let\@oddfoot\@empty\let\@evenfoot\@empty
    \def\@evenhead{\slshape\leftmark\hfil}
    \def\@oddhead{\hfil{\slshape\rightmark}}
    \let\@mkboth\@gobbletwo
    \let\sectionmark\@gobble
    \let\subsectionmark\@gobble
    }
\font\tenrm=cmr10
\font\tenit=cmti10 
\font\tenbf=cmbx10
\font\bfit=cmbxti10 at 10pt
\font\ninerm=cmr9
\font\nineit=cmti9
\font\ninebf=cmbx9
\font\eightrm=cmr8

\def\qed{\hbox{${\vcenter{\vbox{			
   \hrule height 0.4pt\hbox{\vrule width 0.4pt height 6pt
   \kern5pt\vrule width 0.4pt}\hrule height 0.4pt}}}$}}

\begin{document}
\setlength{\textheight}{7.7truein}  

\thispagestyle{empty}

\markboth{\protect{\footnotesize\it Author's Names}}
{\protect{\footnotesize\it Instructions for 
Typesetting Manuscripts (Paper's Title)}}

\normalsize\textlineskip

\setcounter{page}{1}

\copyrightheading{}		

\vspace*{0.88truein}

\centerline{\bf CASIMIR PROBLEM IN SPHERICAL DIELECTRICS:  A QUANTUM 
STATISTICAL MECHANICAL APPROACH }
\baselineskip=13pt
\centerline{}
\centerline{\footnotesize I. BREVIK\footnote{Email address: iver.h.brevik@mtf.ntnu.no}  ~
and J. B. AARSETH\footnote{Email address:  jan.b.aarseth@mtf.ntnu.no.}
.}
\baselineskip=12pt
\centerline{\footnotesize\it Division of Applied Mechanics, Norwegian University of Science and Technology
, N-7491 Trondheim, Norway}
\baselineskip=13pt
\bigskip

\centerline{\footnotesize J. S. H{\O}YE\footnote{Email address:  johan.hoye@phys.ntnu.no.}}

\centerline{\footnotesize\it Department of Physics, Norwegian University of Science and Technology,
N-7491 Trondheim, Norway}
\baselineskip=13pt
\bigskip

\vspace*{0.1truein}
\publisher{Received (Day Month Year)}{Revised (Day Month Year)}

\vspace*{0.21truein}
\abstracts{The Casimir mutual free energy $F$ for a system of two dielectric  concentric nonmagnetic spherical bodies is calculated, at arbitrary temperatures. Whereas $F$ has recently been evaluated for the special case of metals (refractive index $n=\infty$), here analogous results are presented for dielectrics, and shown graphically when $n=2.0$. Our calculational method relies upon quantum statistical mechanics. The Debye expansions for the Riccati-Bessel functions when carried out to a high order are found to be very useful in practice (thereby overflow/underflow problems are easily avoided), and also to give accurate results even for the lowest values of $l$. Another virtue of the Debye expansions is that the limiting case of metals becomes quite amenable to an analytical treatment in spherical geometry. We first discuss the zero-frequency TE mode problem from a mathematical viewpoint and then, as  physical input, invoke the actual dispersion relations. The result of our analysis, based upon  adoption of the Drude dispersion relation as the most correct one at low frequencies, is that the zero-frequency TE mode does $not$ contribute for a metal. Accordingly, $F$ turns out in this case to be only one half of the conventional value. 
}{}{}

\vspace*{10pt}


\vspace*{1pt}\textlineskip	
\section{Introduction}	
\vspace*{-0.5pt}
\noindent

Consider the free energy $F(T)$ at temperature $T$ due to the mutual interaction between two spherical dielectric bodies with concentric surfaces at $r=a$ and $r=b$. Between the two media there is a vacuum.  An analysis of this problem was recently given in Ref.~1, using both quantum statistical methods and field theoretical methods. Whereas the general formalism worked out in Ref.~1  was valid for arbitrary values of the (equal) permittivities $\varepsilon$ in the two dielectric regions, $r<a$ and $r>b$, the explicit evaluations of $F(T)$ for various temperatures $T$ and widths $d=(b-a)$ in Ref.~1 were assuming perfectly conducting walls only, corresponding to $\varepsilon \rightarrow \infty$. Our purpose in the present paper is to extend this calculation so as to include general values of the permittivity. As to our knowledge, such a calculation has not been undertaken before, although there are similarities with the theory given by Kleinert some years ago \cite{kleinert89}. We will assume, as in Ref.~1, that the two media are nonmagnetic. The present account is a short version of a forthcoming paper \cite{brevik01}.

The general conclusion drawn in Ref.~1 was that it is the quantum statistical method that is the most simple and powerful approach when one is to handle the case of general $\varepsilon$. The most central formula in our context is the statistically derived Eq.~(40) in Ref.~1; it gives the value of $\beta F \equiv F/T$ for arbitrary values of temperature, width, and $\varepsilon$. Whereas this equation in terms of a very compact notation, it will be convenient here to rewrite it slightly. Let $m \in \langle -\infty, \infty \rangle$ be an integer corresponding to Matsubara frequencies $K=2\pi m/\beta$; let $n=\sqrt \varepsilon$ be the refractive index of the two media lying at $r<a$ and $r>b$, and let  $s_l(x),~e_l(x)$ be Riccati-Bessel functions with imaginary argument defined according to
\begin{equation}
s_l(x)=\sqrt{\frac{\pi x}{2}}I_\nu(x),~~~~e_l(x)=\sqrt{\frac{2x}{\pi}}K_\nu(x),
\end{equation}
\label{1}
so that their Wronskian becomes $W\{s_l, e_l \}=-1$. Here $\nu=l+1/2$, and $I_\nu,~K_\nu$ are modified Bessel functions. We write the formula as
\begin{equation}
\beta F={\sum_{m=0}^\infty}' \sum_{l=1}^\infty (2l+1)[\ln (1-\lambda_l^{TM})+
\ln(1-\lambda_l^{TE}) ],
\end{equation}
\label{2}
where the prime on the summation sign means that the $m=0$ term is taken with half weight. The two eigenvalues $\lambda_l^{TM}$ and $\lambda_l^{TE}$ in Eq.~(2) correspond to the transverse magnetic and the transverse electric modes. (In the notation of Ref.~1,  $~ \lambda_{\varepsilon l} \equiv \lambda_l^{TM},~\lambda_l \equiv \lambda_l^{TE}$ .) For later use we will write these eigenvalues as ratios. First,
\begin{equation}
\lambda_l^{TM}=\frac{f_1 f_2}{f_3 f_4},
\end{equation}
\label{3}
where
\[ f_1= ns_l'(x)s_l(nx)-s_l(x)s_l'(nx) ,\]
\[ f_2= ne_l'(y)e_l(ny)-e_l(y)e_l'(ny), \]
\[ f_3= ne_l'(x)s_l(nx)-e_l(x)s_l'(nx), \]
\begin{equation}
f_4= ne_l(ny)s_l'(y)-e_l'(ny)s_l(y),
\end{equation}
\label{4}
$x$ and $y$ being the nondimensional frequencies
\begin{equation}
x=2\pi ma/\beta,~~~~y=2\pi mb/\beta.
\end{equation}
\label{5}
We put $\hbar =c=k_B =1$. It should be emphasized that, in contradistinction to the formalism in Ref.~1, the primes in Eqs.~(4) mean derivatives with respect to the {\it whole} argument. Our present way of writing is  in accordance with current usage.

Next, the TE eigenvalues are written as
\begin{equation}
\lambda_l^{TE}=\frac{g_1 g_2}{g_3 g_4},
\end{equation}
\label{6}
where
\[ g_1=s_l'(x)s_l(nx)-ns_l(x)s_l'(nx), \]
\[ g_2= e_l'(y)e_l(ny)-ne_l(y)e_l'(ny), \]
\[ g_3= e_l'(x)s_l(nx)-ne_l(x)s_l'(nx), \]
\begin{equation}
 g_4= e_l(ny)s_l'(y)-ne_l'(ny)s_l(y).
\end{equation}
\label{7}
The following point should be noted. The formulas (4) and (7) contain, as a special case, the situation when the walls at $r=a,b$ are perfectly conducting. This case, as already mentioned, corresponds to setting $n=\sqrt \varepsilon =\infty$. When considering the contribution from zero Matsubara frequency, $m=0$, we are confronted with a delicate two-limit problem. The conventional way to proceed when handling this problem within the framework of nondispersive theory, has been to take the limits in the following order: (i)  Set first $\varepsilon = \infty$; (ii)  take then the limit $m\rightarrow 0$. This is made analytically by observing the  small-argument expressions for the Riccati-Bessel functions.
This way of taking the limits was advocated already in the 1978 paper of Schwinger, DeRaad, and Milton \cite{schwinger78}, and the same procedure was followed in  Ref.~1. If we follow the same procedure also now we get, by insertion into Eq.~(2), the following free energy expression for perfectly conducting walls:
\begin{equation}
\beta F(\varepsilon \rightarrow \infty)={\sum_{m=0}^{\infty}}'\sum_{l=1}^{\infty}
(2l+1)\ln \left\{ \left[ 1-\frac{s_l(x)}{e_l(x)}\frac{e_l(y)}{s_l(y)}\right]
\left[ 1-\frac{s_l'(x)}{e_l'(x)}\frac{e_l'(y)}{s_l'(y)}\right] \right\},
\end{equation}
\label{8}
which is in agreement with Eq.~(68) in Ref.~1. If we next, following the same recipe, let $x\rightarrow 0,~y\rightarrow 0$, we obtain the following contribution from $m=0$:
\begin{equation}
\beta F^{conv}(\varepsilon \rightarrow \infty, m=0)=\sum_{l=1}^\infty (2l+1)\ln \left[ 1-\left( \frac{a}{b}\right)^{2l+1} \right],
\end{equation}
\label{9}
again in agreement with Ref.~1, Eq.~(79). This is the conventional result. {\it Both} the two electromagnetic modes are in this way found to contribute equally to the sum in Eq.~(9).

A discussion has recently arisen as to whether this recipe for dealing with the $m=0$ term in the TE mode is really correct. The problem becomes most acute in the high $T$ regime, but is present at moderate and low temperatures also. We refer here to the paper of Bostr\"{o}m and Sernelius \cite{bostrom00}, questioning this point, and the subsequent comment of Lamoreaux \cite{lamoreaux01}. What has been most welcome in recent years are the accurate experiments on the Casimir force, due to  Lamoreaux \cite{lamoreaux97} and Mohideen {\it et al.} \cite{mohideen98} \cite{roy99} \cite{harris00}. By means of these experiments it becomes much easier to formulate a sound theory. Several theoretical papers have lately appeared, discussing the experiments' various facets \cite{lamoreaux98} \cite{lambrecht00} \cite{svetovoy00}  \cite{bordag00} \cite{klimchitskaya01}. An extensive recent review has been given by Bordag {\it et al.} \cite{bordag01}. We also mention several other related papers \cite{milton99} \cite{barton01} \cite{feinberg01}, of a more general nature; also these  being concerned with finite temperature effects in a Casimir context.

Below, we will make use of the present formalism to analyze how the $m=0$ case works out for the spherical geometry. It turns out that the formalism becomes quite amenable. An important ingredient in our analysis is the use of the Debye expansions for the Riccati-Bessel functions. They make the formalism quite transparent, and they imply that the overflow or underflow problems that so easily turn up in this sort of calculations, are easily abandoned. Moreover, when carrying out the Debye expansion to a high order (18th order in the quantity $\theta$ defined by Eq.~(15) below), the numerical accuracy becomes high for all values of $l$, quite sufficient for all practical calculations. We analyze the problem of metals first from a mathematical point of view and then, inserting the plasma dielectric model versus the Drude model as physical input, show how the result for $F$ depends critically on which dispersion relation one chooses. As physically the Drude model is preferable at very low frequencies, we conclude that the $m=0$ TE mode does {\it not} contribute to $F$ for a metal. This implies that the conventional expression for $F$ for a metal has to be multiplied by one half. This is also in agreement with our earlier statistical mechanical considerations for the static case in Sec.~III in Ref.~1.

\section{Numerical Evaluation}
\noindent

\subsection{\bfit The Debye expansions}
 We define the nondimensional temperature:
\begin{equation}
t=\frac{2\pi a}{\beta},
\end{equation}
\label{10}
implying $x=mt$, and write the Debye expansions of the Riccati-Bessel functions in the form \cite{abramowitz72}
\begin{equation}
s_l(x)=\frac{1}{2}\frac{\sqrt{z(x)}}{[1+z^2(x)]^{1/4}}\, e^{\nu \eta(x)}\,A[\theta(x)],
\end{equation}
\label{11}
\begin{equation}
e_l(x)=\frac{\sqrt{z(x)}}{[1+z^2(x)]^{1/4}}\,e^{-\nu \eta(x)}\,B[\theta(x)],
\end{equation}
\label{12}
\begin{equation}
s_l'(x)=\frac{1}{2}\frac{[1+z^2(x)]^{1/4}}{\sqrt{z(x)}}\,e^{\nu \eta(x)}\,C[\theta(x)],
\end{equation}
\label{13}
\begin{equation}
e_l'(x)=-\frac{[1+z^2(x)]^{1/4}}{\sqrt{z(x)}}\,e^{-\nu \eta(x)}\,D[\theta(x)].
\end{equation}
\label{14}
Here $\nu=l+1/2,~l=1,2,...$ and
\[ z(x)=x/\nu,~~~~\theta(x)=[1+z^2(x)]^{-1/2}, \]
\begin{equation}
\eta(x)=\frac{1}{\theta(x)}+\ln \frac{z(x)}{1+1/\theta(x)} 
\end{equation}
\label{15}
($\theta$ is the same as the symbol $t$ in Ref.~20). The four polynomials, $A(\theta), B(\theta), C(\theta), D(\theta)$, are found to be of order unity. In Ref.~21 we expanded them to order $\theta^{18}$. These expansions, which will not be reproduced here, are found to be easily handled by a computer. The polynomials possess the following important property:
\begin{equation}
\{ A(\theta), B(\theta), C(\theta), D(\theta) \} \rightarrow 1 ~~~{\rm when}~~~ \theta \rightarrow 0.
\end{equation}
\label{16}
The factors in Eqs.~(11)-(14) that can take extreme values, are the exponentials. They are easily dealt with analytically in the Debye formalism. We avoid in this way the overflow/underflow problems that might easily occur from a simple use of the computer Bessel library.

It is now convenient to calculate the following ratios between the functions defined in Eq.~(4):
\begin{equation}
\frac{f_1}{f_3}=-\frac{1}{2}e^{2\nu \eta(x)}\,\frac{n^2\gamma C[\theta(x)]-A[\theta(x)]C[\theta(nx)]/A[\theta(nx)]}
{n^2\gamma D[\theta(x)]+B[\theta(x)]C[\theta(nx)]/A[\theta(nx)]}
 \end{equation}
\label{17}
and
\begin{equation}
\frac{f_2}{f_4}=-2e^{-2\nu \eta(y)}\, \frac{n^2\delta D[\theta(y)]-B[\theta(y)]D[\theta(ny)]/B[\theta(ny)]}
{n^2\delta C[\theta(y)]+A[\theta(y)]D[\theta(ny)]/B[\theta(ny)]},
\end{equation}
\label{18}
where $\gamma$ and $\delta$ are the coefficients
\begin{equation}
\gamma= \sqrt{\frac{1+z^2(x)}{1+z^2(nx)}},~~~~\delta=\sqrt{\frac{1+z^2(y)}{1+z^2(ny)}}.
\end{equation}
\label{19}
Consequently (cf. Eq.~(3))
\begin{equation}
\lambda_l^{TM}=e^{2\nu [\eta(x)-\eta(y)]}\, \times [...],
\end{equation}
\label{20}
where
\begin{equation}
\eta(x)-\eta(y)=\sqrt{1+z^2(x)}-\sqrt{1+z^2(y)}+\ln \left( \frac{a}{b}\frac{1+\sqrt{1+z^2(y)}}
{1+\sqrt{1+z^2(x)}} \right),
\end{equation}
\label{21}
and where [...] is the polynomial ratio following from Eqs.~(17) and (18).

Similarly
\begin{equation}
\frac{g_1}{g_3}=-\frac{1}{2}e^{-2\nu \eta(x)}\,\frac{\gamma C[\theta(x)]-A[\theta(x)]C[\theta(nx)]/A[\theta(nx)]}
{\gamma D[\theta(x)]+B[\theta(x)]C[\theta(nx)]/A[\theta(nx)]},
\end{equation}
\label{22}
\begin{equation}
\frac{g_2}{g_4}=-2e^{-2\nu \eta(y)}\, \frac{\delta D[\theta(y)]-B[\theta(y)]D[\theta(ny)]/B[\theta(ny)]}
{\delta C[\theta(y)]+A[\theta(y)]D[\theta(ny)]/B[\theta(ny)]},
\end{equation}
\label{23}
so that $\lambda_l^{TE}$ takes the same form as $\lambda_l^{TM}$, Eq.~(20), only with the difference that [...] now is formed from the polynomials in Eqs.~(22) and (23).

\subsection{\bfit Numerical results for dielectrics}

\begin{figure}[htb]
\centerline{\psfig{file=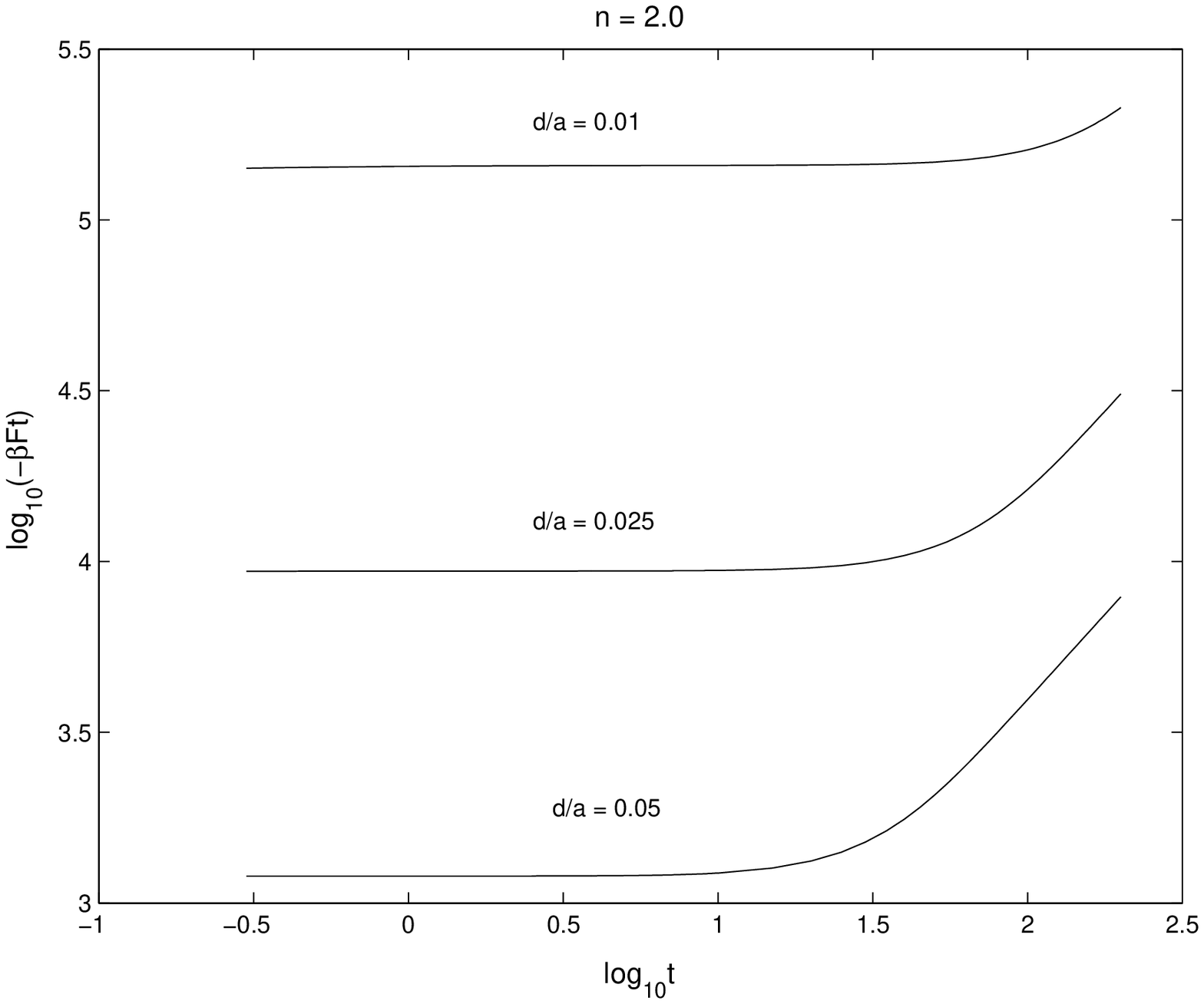,width=12cm,height=8cm,angle=0}}
\vspace*{12pt}

\bigskip

\bigskip

\fcaption{Logarithm of nondimensional free energy, $\log_{10}(-\beta Ft)$, versus relative width $d/a=(b-a)/a$ for various temperatures $t=2\pi a/\beta$. Refractive index $n=2.0.$}
\end{figure}

Figure~1 shows, as an example, how $\log_{10}(-\beta Ft)$ varies with $\log_{10}t$ when $n=2.0$. It shows clearly the presence of a low-temperature plateau. For higher values of $t$, there is a gradual change into the region where $F$ varies linearly with $t$. 
Our calculations show that the magnitude $|F|$ of the free energy for an ordinary dielectric is much less than for a metal ($n=\infty$)\cite{hoye01}. This is as we would expect.

Generally, we found the asymptotic Debye expansions to be useful for $x>10$ and/or $l>9$. Then, an accuracy of 8 digits for the individual terms was achieved. Below these limits for $x$ and $l$, we employed the machine-generated Bessel functions. For small values of $d/a$ and $t$, slow convergence was observed. The summation of the series thus became rather demanding.  As an example, when $d/a=0.05,~t=0.01$, about 1.1 million terms were needed, if we truncated the summation at  $\varepsilon =10^{-9}$ (here $\varepsilon$ means the ratio between a general term in the series and the sum). The sum itself is however accurate only up to 4 or 5 digits.

An important result was that even for low values of $l$, the asymptotic series gave very good results. One reason for this is the high-order expansions used for the polynomials $A,B,C,D$.  It seems most likely that the Debye expansions (at least when carried out to order $\theta^{18}$) can be used for {\it all} $x$ and $l$, for all practical purposes.

\section{The Limiting Case of a Metal}

\subsection{\bfit The nondispersive case}

It turns out that the Debye expansions are quite useful also for the purpose of analyzing metals. We first assume a nondispersive medium, so that the case of metals corresponds to $\varepsilon \rightarrow \infty$, or $n\rightarrow \infty$.

For simplicity let us categorize how to take the two limits, i.e. the limit on $n$, and the limit on the Matsubara number $m$. The first option, which we shall call option A, means taking {\it first} $n\rightarrow \infty$, {\it thereafter} $m\rightarrow 0$. Option B reverses the succession of $n$ and $m$.

We consider the TM mode, first employing option A. Taking the first limit $n\rightarrow \infty$, it follows that $\theta(x)$ is finite, whereas $\theta(nx)\rightarrow 0$. Thus all polynomials $\{A,B,C,D\}[\theta(x)]$ taken at argument $\theta(x)$ are finite, whereas $\{A,B,C,D\}[\theta(nx)]\rightarrow 1$ according to Eq.~(16). Since $n^2\gamma$ and $n^2\delta$ are proportional to $n$ for large $n$ according to Eq.~(19), we get from Eqs~(17)-(20)
\begin{equation}
\lambda_l^{TM}(n\rightarrow \infty)=e^{2\nu [\eta(x)-\eta(y)]}\,\frac{C[\theta(x)]D[\theta(y)]}
{D[\theta(x)]C[\theta(y)]}.
\end{equation}
\label{24}
Taking the limit $m\rightarrow 0$ we have from Eq.~(21) $\eta(x)-\eta(y)\rightarrow \ln(a/b)$, so that the $m=0$ free energy becomes
\begin{equation}
\beta F^{TM}(m=0)=\frac{1}{2}\sum_{l=1}^\infty (2l+1)\ln \left[1-\left(\frac{a}{b}\right)^{2l+1} \right].
\end{equation}
\label{25}
Consider next option B. When $m\rightarrow 0$  we have $\theta(x)\rightarrow 1,~\theta(nx)\rightarrow 0$, implying that $\gamma \rightarrow 1,~\delta \rightarrow 1$ and $\lambda_l^{TM}(m=0)=(a/b)^{2l+1}$. Thus we obtain the same expression for the $m=0$ TM free energy as before, Eq.~(25). The robustness of the TM calculated free energy is actually what we could expect on physical grounds: the TM mode means that the magnetic field is transverse to the radius vector ${\bf r}$, thus parallel to the spherical surfaces at $r=a,b$. This is precisely the natural electromagnetic boundary condition for the magnetic field at perfect conducting surfaces.

Consider then the TE mode. Employing option A we get
\begin{equation}
\lambda_l^{TE}(n\rightarrow \infty)= e^{2\nu [\eta(x)-\eta(y)]}\; \lim_{n\rightarrow \infty} \frac 
{\{\gamma C[\theta(x)]-A[\theta(x)]\}\{\delta D[\theta(y)]-B[\theta(y)]\} }
{\{\gamma D[\theta(x)]+B[\theta(x)]\}\{\delta C[\theta(y)]+A[\theta(y)]\} }.
\end{equation}
\label{26}
The difference between this case and the preceding case lies in the sensitivity of Eq.~(26) with respect to $\gamma$ and $\delta$. From Eq.~(19) it follows that $\gamma\rightarrow 0,~ \delta\rightarrow 0$ implying that, when we take the limit $m\rightarrow 0$, $\lambda_l^{TE} \rightarrow (a/b)^{2l+1}$. Then, Eq.~(2) shows that the TE contribution to the $m=0$ free energy becomes the same as Eq.~(25). 

Consider now option B. We obtain $\gamma \rightarrow 1,~\delta\rightarrow 1$. Then, according to Eqs.~(22) and (23),  $~\lambda_l^{TE}\rightarrow 0$ when $m\rightarrow 0$. Consequently
\begin{equation}
B:~~~\beta F^{TE}(m=0)=0.
\end{equation}
\label{27}
Option B thus gives only half as large total free energy as the conventional result, Eq.~(9). It is worth noticing that option B is in accordance with the quantum statistical mechanical result for the static mode \cite{hoye01}.

In order to decide between the two options we have to bring physics into the consideration, i.e. the appropriate dispersion relation. This is the topic of the next subsection.

\subsection{\bfit The dispersive case}

Let $\hat{\omega}$ be the frequency along the imaginary frequency axis. There are essentially two dispersion relations on the market. The first is the plasma relation
\begin{equation}
\varepsilon(i\hat{\omega})=1+\frac{\omega_p^2}{\hat{\omega}^2},
\end{equation}
\label{28}
which is valid at the far ultraviolet for the light elements and at the X-ray region for heavier elements (Sec. 78 in Ref.~22). If we nevertheless employ Eq.~(28) even near $\hat\omega=0$, it follows that  $n(i\hat{\omega})\hat{\omega}a/\nu \rightarrow x_p/\nu$ where, in dimensional units, $x_p \equiv \omega_p a/c$. Taking typically $\omega_p \sim 3\times 10^{16}~{\rm s}^{-1}$ and $a \sim 1$ cm we get $x_p \sim 10^6$. In practice, the most significant values of $l$ are much lower than this.  We can thus assume that $x_p/\nu \gg 1$, so that in practice $\gamma \rightarrow 0,~\delta \rightarrow 0$. That is, we recover in this way option A, and thereby the conventional result,  Eq.~(9).

Consider next the {\it Drude model} for the dielectric, corresponding to
\begin{equation}
\varepsilon(i\hat{\omega})=1+\frac{\omega_p^2}{\hat{\omega}(\hat{\omega}+\gamma)},
\end{equation}
\label{29}
$\gamma$ being the relaxation frequency. According to this relation $n(i\hat{\omega})\hat{\omega} \rightarrow 0$ when $\hat{\omega}\rightarrow 0$, implying that $\gamma \rightarrow 1,~\delta \rightarrow 1$. That is, we recover  option B. The total $m=0$ free energy for a metal is thus according to the Drude model predicted to be one half of the conventional expression (9).

Which of the two dispersion relations is correct? In our opinion, it is the relation (29), when $\hat{\omega}\rightarrow 0$.  On physical grounds the permittivity has to be inversely proportional to the frequency at low frequencies; cf. Sec. 77 in Ref.~22. Explicitly,  $\varepsilon(i\hat{\omega})=\sigma/\hat{\omega}$, where $\sigma$ is the conductivity. This is a result following directly from Maxwell's equations. The Drude model satifies this requirement. Thus both the Drude model (and, as we have seen, statistical mechanical methods), support the option B above. The plasma model, Eq.(28), as we have noted, is appropriate only at higher frequencies.

\section{Conclusions, and Final Remarks}

We may summarize as follows:

1.  For a nondispersive dielectric, the formalism in spherical geometry becomes quite tractable. The free energy at finite temperatures is calculated from Eq.~(2). Employing the Debye expansions up to 18th order in the quantity $\theta$ (cf. Eq.~(15)), good accuracy is achieved for all values of $x$ and $l$.

2.  In the special case of a metal, adopting the Drude dispersion relation, we find that the $m=0$ TE mode does not contribute. The total $m=0$ free energy for a metal becomes accordingly only one one half of the conventional expression (9).

3.  As an extension of the above considerations, one may inquire about the magnitude of the $m=0$ contribution to the free energy for a dielectric. This case is treated in more detail in Ref.~3.  From Eq.~(2) one has, for arbitrary temperatures,
\begin{equation}
\beta F(m=0)=\frac{1}{2}\sum_{l=1}^\infty (2l+1)\ln (1-\lambda_l^{TM})
\end{equation}
\label{30}
($\lambda_l^{TE}$  does not contribute for a dielectric). We may define $Y$ as the ratio between  $F(m=0)$ and the expression (2) for the full free energy:
\begin{equation}
Y=\frac{F(m=0)}{F}.
\end{equation}
\label{31} 
For given $d/a$, $Y$ thus becomes a function only of $t$. In Ref.~3, we show how $Y$ varies with $t$ for various values of $d/a$, for a fixed value of $n$.

One general conclusion to be drawn from this calculation is that the less the value of $d/a$, the less becomes the importance of the $m=0$ term. This is a result that can be understood physically: when the slit is narrow, we can approximately regard the system as a conventional two-plates system. For the latter geometry, it is  known that the classicality condition can be written as $dT \gg 1$, where $d$ is the distance between the plates (cf. Sec. 82 in Ref.~23). When $d$ decreases the system thus becomes more and more a quantum-mechanical system, necessitating an increasing large region of frequencies determining the value of $F$. The relative importance of the low frequencies, in particular that of $m=0$, thus has to diminish, in accordance with the result of the calculation.

\nonumsection{Acknowledgments}
\noindent
I thank Michael Bordag and his group for arranging a pleasant and stimulating Workshop in Leipzig. Financial support from NTNU (Trondheim) is acknowledged.

\nonumsection{References}


\vspace*{-8pt}

\begin{thebibliography}{000}


\bibitem{hoye01}
J. S. H{\o}ye, I. Brevik, and J. B. Aarseth,  {\nineit Phys. Rev. E} {\ninebf 63}, 051101 (2001).

\bibitem{kleinert89}
H. Kleinert, {\nineit Phys. Lett. A} {\ninebf 136}, 253 (1989).

\bibitem{brevik01}
I. Brevik, J. B. Aarseth, and J. S. H{\o}ye, in preparation.

\bibitem{schwinger78}
J. Schwinger, L. L. DeRaad, Jr., and K. A. Milton, {\nineit Ann. Phys. (N.Y.)} {\ninebf 115}, 1 (1978).

\bibitem{bostrom00}
M. Bostr\"{o}m and Bo E. Sernelius, {\nineit Phys. Rev. Lett.} {\ninebf 84}, 4757 (2000).

\bibitem{lamoreaux01}
S. K. Lamoreaux, quant-ph/0007029.

\bibitem{lamoreaux97}
S. K. Lamoreaux, {\nineit Phys. Rev. Lett.} {\ninebf 78}, 5 (1997).

\bibitem{mohideen98}
U. Mohideen and A. Roy, {\nineit Phys. Rev. Lett.} {\ninebf 81}, 4549 (1998).

\bibitem{roy99}
A. Roy, C.-Y. Lin, and U. Mohideen, {\nineit Phys. Rev. D} {\ninebf 60}, 111101(R) (1999).

\bibitem{harris00}
B. W. Harris, F. Chen, and U. Mohideen, {\nineit Phys. Rev. A} {\ninebf 62}, 052109 (2000).

\bibitem{lamoreaux98}
S. K. Lamoreaux, {\nineit Phys. Rev. Lett.} {\ninebf bf 81}, 5475 (1998); {\nineit Phys. Rev. A} {\ninebf 59}, R3149 (1999).

\bibitem{lambrecht00}
A. Lambrecht and S. Reynaud, {\nineit Eur. Phys. J. D} {\ninebf 8}, 309 (2000) (quant-ph/9907105); {\nineit Phys. Rev. Lett.} {\ninebf bf 84}, 5672 (2000).

\bibitem{svetovoy00}
V. B. Svetovoy and M. V. Lokhanin, {\nineit Mod. Phys. Lett. A} {\ninebf 15}, 1437 (2000) (quant-ph/0008074); quant-ph/0004004; {\nineit Phys. Lett. A} {\ninebf 280}, 177 (2001) (quant-ph/0101124).

\bibitem{bordag00}
M. Bordag, B. Geyer, G. L. Klimchitskaya, and V. M. Mostepanenko, {\it Phys. Rev. Lett.} {\ninebf bf 85}, 503 (2000).

\bibitem{klimchitskaya01}
G. L. Klimchitskaya and V. M. Mostepanenko, {\it Phys. Rev. A} {\ninebf 63}, 062108 (2001).

\bibitem{bordag01}
M. Bordag, U. Mohideen, and V. M. Mostepanenko, {\nineit Phys. Reports} {\ninebf 353}, 1 (2001) (quant-ph/0106045).

\bibitem{milton99}
K. A. Milton, in {\nineit Applied Field Theory, Proceedings of the 17th Symposium on Theoretical Physics}, edited by C. Lee, H. Min, and Q.-H. Park (Chungbum, Seoul, 1999) (hep-th/9901011).

\bibitem{barton01}
G. Barton, {\nineit Phys. Rev. A} {\ninebf bf 64}, 032103 (2001).

\bibitem{feinberg01}
J. Feinberg, A. Mann, and M. Revzen, {\nineit Ann. Phys. (N.Y.)} {\nineit \bf 288}, 103 (2001).

\bibitem{abramowitz72}
M. Abramowitz and I. A. Stegun, {\nineit Handbook of Mathematical Functions} (Dover, New York, 1972).

\bibitem{brevik87}
I. Brevik, {\nineit J. Phys. A: Math. Gen.} {\ninebf 20}, 5189 (1987).

\bibitem{landau84}
L. D. Landau and E. M. Lifshitz, {\nineit  Electrodynamics of Continuous Media}, 2nd ed. (Pergamon Press, Oxford, 1984).

\bibitem{landau80}
E. M. Lifshitz and L. P. Pitaevskii, {\it Statistical Physics, Part 2} (Pergamon Press, Oxford, 1980). 



\end{thebibliography}
\end{document}